\documentclass[aps,article,reprint,amsmath,amsfonts]{revtex4-1} 
\usepackage{amssymb, bm, hyperref}
\usepackage{graphicx}
\pdfoutput=1

\begin{document}

\title{Holey topological thermoelectrics}

\author{O.~A.~Tretiakov, Ar. Abanov, and Jairo Sinova}

\affiliation{Department of Physics,
	    Texas A\&M University,
            College Station, TX 77843, USA}

\date{June 30, 2011}

\begin{abstract}
We study the thermoelectric properties of three-dimensional
topological insulators with many holes (or pores) in the bulk. We show
that at high density of these holes the thermoelectric figure of merit
$ZT$ can be large due to the contribution of the conducting surfaces and
the suppressed phonon thermal conductivity. The maximum efficiency can
be tuned by an induced gap in the surface states dispersion through
tunneling or external magnetic fields. The large values of $ZT$, much
higher than unity for reasonable parameters, make this system a strong
candidate for applications in heat management of nanodevices,
especially at low temperatures.
\end{abstract}

\maketitle

Efficient heat management in nanodevices and energy conversion from
wasted heat have been some of the key driving motivations for the
search of new thermoelectric materials \cite{Snyder2008, Tritt99,
  Dresselhaus93, mukerjee07, Markussen09, Murakami2010, Ghaemi10,
  TretiakovAPL10, Ho-KiLyeo04, venkatasubramanian01, zhang:062107,
  TeweldebrhanAPL10, Teweldebrhan10, DiVentra11}.  Their efficiency
rests on the ability of a material to have a low thermal conductivity
and high thermopower and electric conductivity. The large thermopower
requires steep dependence of the electronic density of states on
energy, which can be achieved by having the chemical potential close
to the bottom of a band.  The relatively high conductivity demands the
gap to be low. These combined requirements point to semiconductors
with heavy elements as the best candidates for the thermoelectric
materials.

A typical band gap for an efficient thermoelectric device at room
temperature should be of the order of few hundred meV.  Such a low gap
can appear in the materials either accidentally or due to a large
enough spin-orbit coupling that leads to the inversion of the band
structure, thus providing a band-gap of the order of the spin-orbit
coupling energy. This latter class of materials, besides being
highlighted by their high thermoelectric efficiency, has been
identified as having an insulating bulk and topologically protected
conducting surfaces with Dirac-like band structure
\cite{TI_physics_today, Fu07, Hsieh2008, Chen2009}; hence their name,
topological insulators (TI).

The bulk thermoelectric properties of this family of semiconductors
have been extensively studied both theoretically \cite{ Murakami2010,
  Ghaemi10, TretiakovAPL10} and experimentally
\cite{venkatasubramanian01, zhang:062107, TeweldebrhanAPL10,
  Teweldebrhan10}.  The rather high thermoelectric efficiency found in
these materials is the property of the band structure 
and the heavy atomic masses, and is unrelated
to the contribution of the protected surface states --- the main focus
of this paper.  As with most thermoelectric materials, the main
stumbling block for increasing the thermoelectric efficiency is the
high phonon thermal conductivity. There have been many attempts to
decrease it while leaving the electric conductivity intact
\cite{Poudel2008, Tang10}, e.g., by introducing disorder in the
material that suppresses phonon transport (phononic glass). The bulk
electronic states are however often affected by this disorder as well
and it is very difficult to reduce thermal conductivity without also
reducing thermopower and electric conductivity \cite{Snyder2008}.

Here we propose a new approach to increase thermoelectric efficiency
by combining the novel properties of the TI materials, {\it i.e.},
their protected surface states, with methods to reduce thermal
conductivity that can enhance the contribution to the conductivity and
thermopower from these topological surface states by increasing the
surface to volume ratio.

The thermopower efficiency is characterized by the dimensionless
thermoelectric figure of merit $ZT = \sigma S^2 T/\kappa$, where
$\sigma$, $S$, $\kappa$, and $T$ are the electric conductivity,
thermopower, thermal conductivity, and temperature, respectively. This
quantity corresponds to the ratio of the output power to the rate of
the heat power consumption under an applied temperature gradient.

\begin{figure}
\includegraphics[width=0.8 \columnwidth]{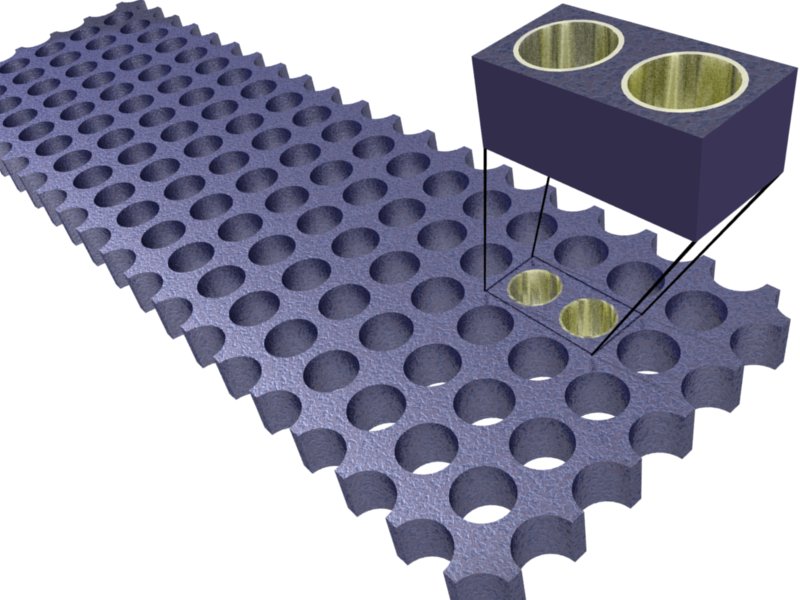} 
\caption{A sample made of a holey 3D topological insulator. These
  holes increase the surface to bulk ratio and reduce its phonon
  thermal conductivity.  The inset shows two such holes with the Dirac
  like metallic surface states.}
\label{fig:holey_sample} 
\end{figure}

The high surface to bulk ratio can be achieved by nanostructuring 3D
topological insulators with many holes in the transport direction.
The surface of each hole behaves as a 2D conductor. The effect of the
holes is threefold: 1) to suppress phonon thermal conductivity -- the
strongest limiting factor for $ZT$, 2) to increase conductivity due to
a high contribution of conducting surface, and 3) to
allow for the surface states to tunnel and create a smaller subgap to
increase $ZT$ further. A sketch of the sample made of a 3D topological
insulator with a high density of holes is depicted in
Fig.~\ref{fig:holey_sample}.  Even at high density of these holes, the
topologically protected surface states remain robust. It has been
shown that these states survive in the films as thin as of three or
more quintuple layers \cite{ParkPRL10}.

Reducing thermal conductivity caused by the increased phonon
scattering off grain boundaries and defects has been experimentally
confirmed recently.  For example, in BiSbTe bulk alloys higher $ZT$
has been observed in experiments with nanocrystalline bulk samples
made by pressing nanopowders at high temperatures \cite{Poudel2008}.
Furthermore, high density of periodic nanoscale holes in silicon thin
films has been shown to reduce the thermal conductivity by two orders
of magnitude \cite{Tang10}.

The unique feature of the topological insulators is that, although
they have a bulk energy gap of $2\Delta_0$, they also have ungaped
surface states. These propagating states are confined to the close
proximity of the surfaces. Their existence is protected by the
topology of the bulk band structure.  These states have cone-like 2D
Dirac spectrum, $E=\pm v\hbar |k|$, where $v$ is the constant Dirac
electron velocity
\footnote{We consider $v$ to be energy independent in a wide range of
  energies}.  Application of a magnetic field or hybridization of
these states due to close proximity of two surfaces can induced a
controllable Dirac subgap $\Delta$ \cite{Linder09, LuPRB10}. Then the
spectrum takes the form $E=\pm \sqrt{v^2\hbar^2
  k^2+\Delta^2}$. Generally this subgap is much smaller than the bulk
gap, $\Delta\ll \Delta_0$. For 3D TIs $2\Delta_0$ is roughly from
$0.15$ eV (for $\rm{Bi}_{2}\rm{Te}_3$) to $0.3$ eV (for
$\rm{Bi}_{2}\rm{Se}_3$).

\begin{figure}
\includegraphics[width=1.0 \columnwidth]{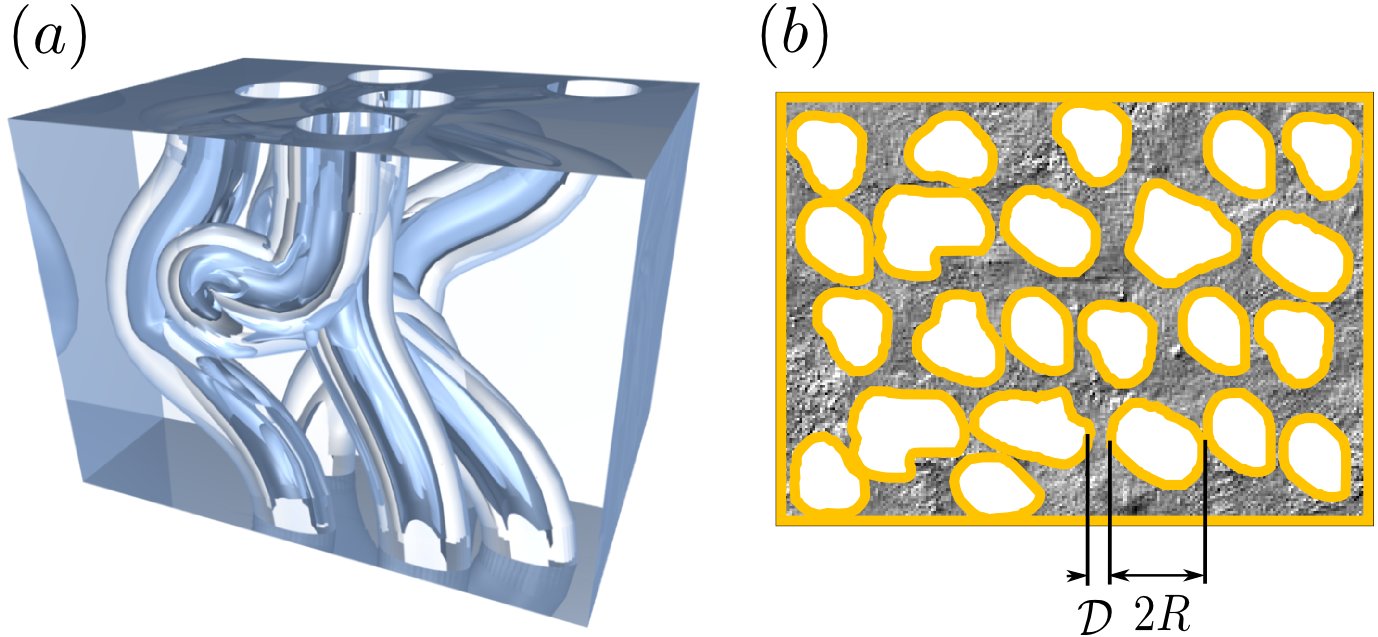} 
\caption{(a) A sample of 3D topological insulator with
  randomly distributed pores. (b) Top view of small part of the sample
  with the holes. ${\cal D}$ represents the average distance between
  randomly distributed pores (or holes).}
\label{fig:porous} 
\end{figure}

To study the transport properties of these materials we use linear
response theory \cite{AshcroftMermin, marder}. The electric ($j^e$)
and thermal ($j^q$) currents are given by linear combinations of the
chemical potential and temperature gradients: $j^e/e= L_0\nabla \mu +
L_1 (\nabla T)/T$ and $j^q= -L_1\nabla \mu - L_2 (\nabla T)/T$, where
$e$ is the electron charge.  Using Onsager relations, one can find
from these equations the electrical conductivity $\sigma= e^2 L_0$,
thermopower $S=-L_1/(eTL_0)$, and electronic thermal conductivity
$\kappa_{e}=(L_0 L_2 -L_1^2)/(TL_0)$. The figure of merit, $ZT$, can
then be represented in terms of these linear coefficients as
\begin{equation}
ZT = \frac{L_{1}^2}{L_{0}(L_{2}+\kappa_{ph}T) -L_{1}^2}.
\label{ZT2}
\end{equation}
Here $\kappa_{ph}$ is the phonon contribution to the thermal
conductivity in the bulk (phonon contribution to the thermal
conductivity for the surface TI states is much smaller than that in
the bulk). In Eq.~(\ref{ZT2}) it is assumed that the transport
coefficients have bulk and surface contributions $L_n = L_{b,n} +
L_{s,n}/{\cal D}$, where ${\cal D}=(A-\sum_n \pi R_n^2)/\sum_n 2\pi
R_n$ is the factor related to surface/bulk ratio (porosity) and has
dimension of length. The parameter ${\cal D}$ characterizes the
average distance between the pores (holes) of the average radius $R$,
see Fig.~\ref{fig:porous} (b).

\begin{figure}
\includegraphics[width=1.0 \columnwidth]{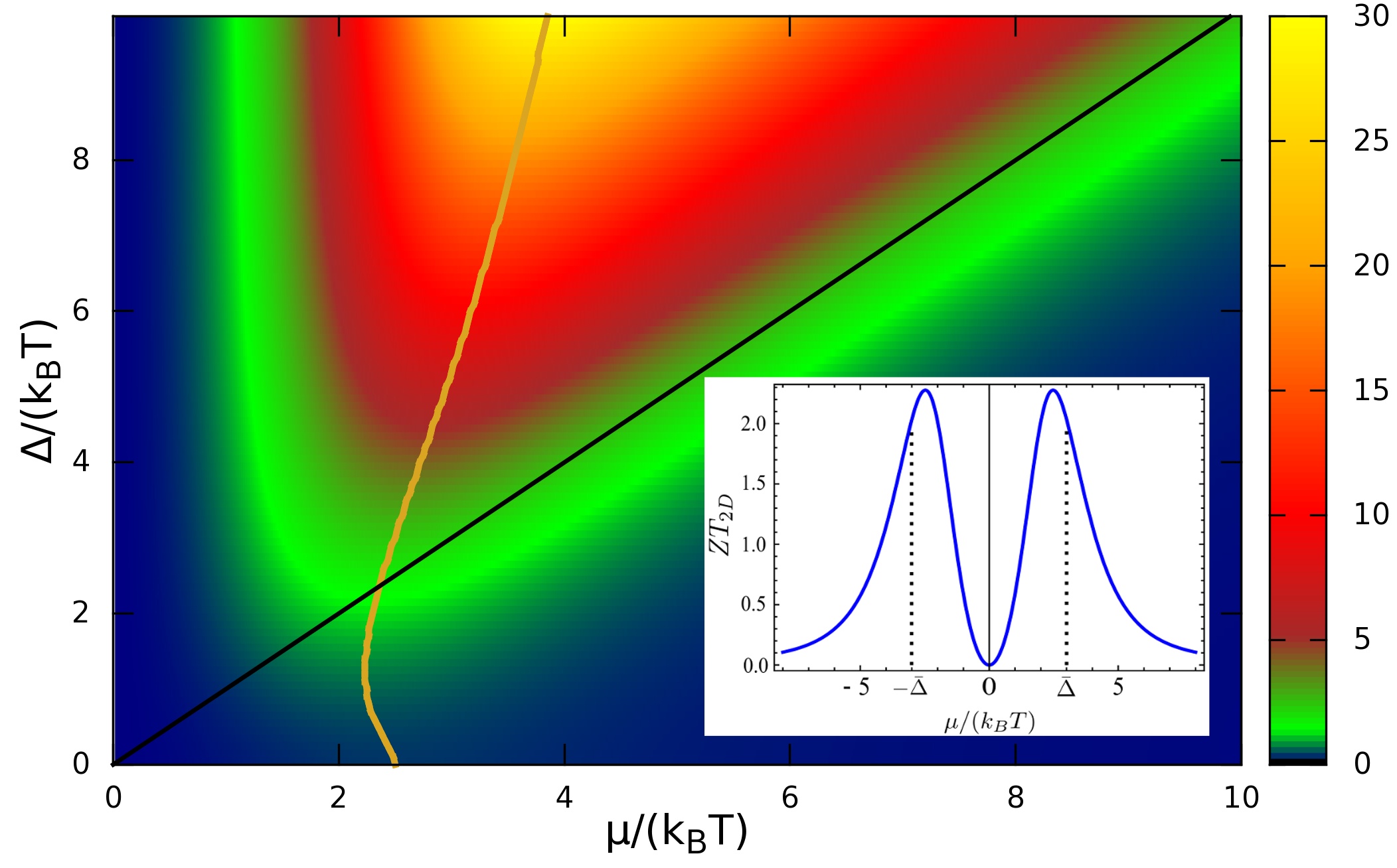} 
\caption{The color plot of $ZT$ only for the surface
  states as a function of the induced subgap $\Delta/(k_BT)$ and
  chemical potential $\mu/(k_BT)$ measured in units of
  temperature. The inset shows $ZT$ calculated only for the surface
  states with the subgap $\Delta/(k_BT)=3$. The solid golden line
  represents the maximum $ZT$ at a fixed $\Delta$. The black line
  corresponds to $\mu_{max} =\Delta$.}
\label{fig:ZT_gap_mu} 
\end{figure}

It follows from Eq.~(\ref{ZT2}) that at small ${\cal D}$, the
contribution to $ZT$ mostly comes from 2D surface states and in this
limit $ZT$ is given by
\begin{equation}
ZT_{2D}=  
\frac{L_{s,1}^2}{L_{s,0}L_{s,2} -L_{s,1}^2}.
\label{ZT_2D}
\end{equation}
$ZT_{2D}$ is shown in the inset of Fig.~\ref{fig:ZT_gap_mu} for $\Delta/(k_B
T)=3$ as a function of the chemical potential. The color plot of
$ZT_{2D}$ as a function of both the induced subgap $\Delta/(k_B T)$ and
chemical potential $\mu/(k_B T)$ is shown in
Fig.~\ref{fig:ZT_gap_mu}. For a fixed induced gap $\Delta/(k_B T)$ we
plot the location of the maximum $ZT$ achievable by tuning the
chemical potential (golden line in Fig.~\ref{fig:ZT_gap_mu}). The very
high values of $ZT$ in reality are not reachable since any small
contribution from the phonon thermal conductivity will reduce $ZT$.

\begin{figure}
\includegraphics[width=1.0 \columnwidth]{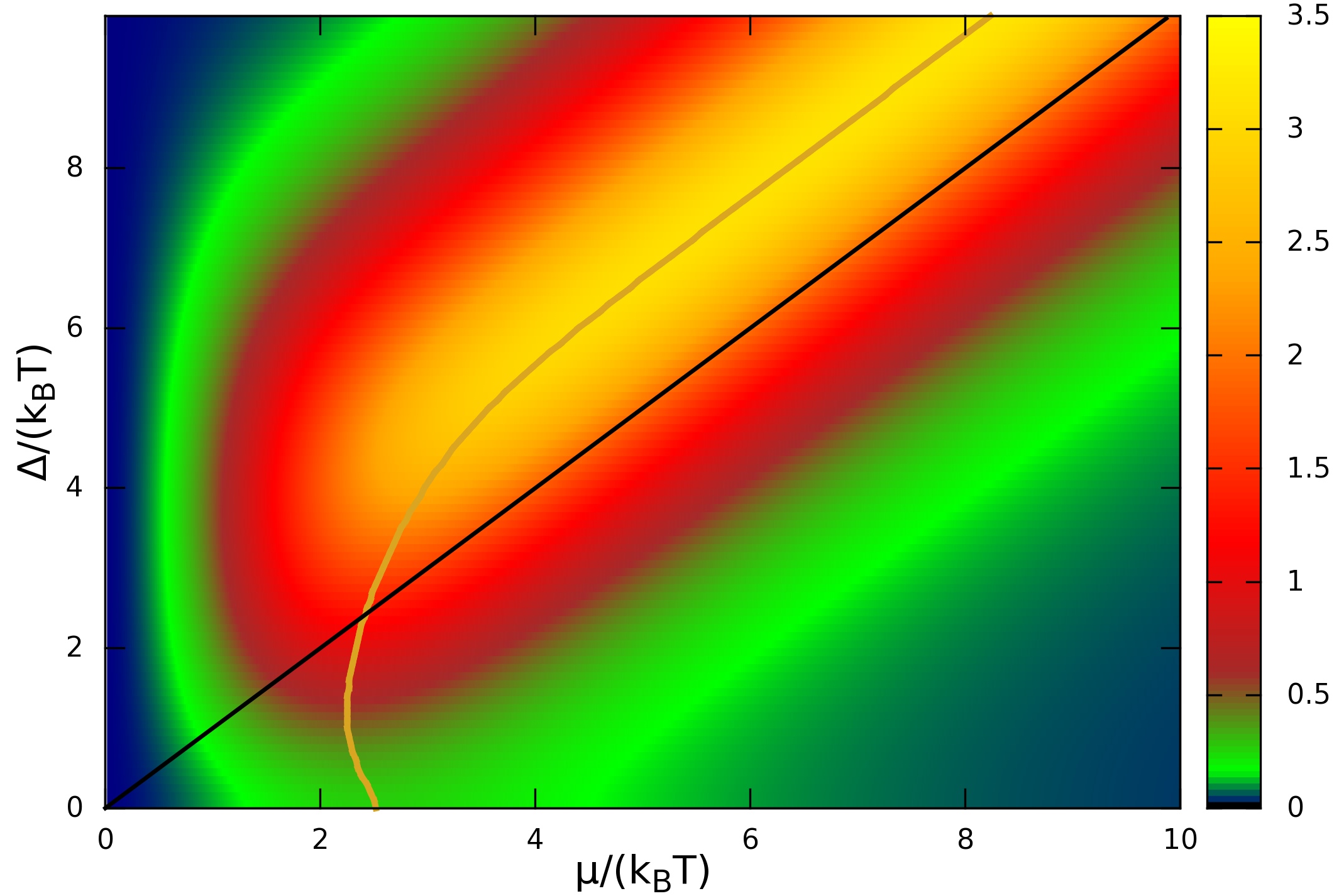} 
\caption{$ZT$ as a function of the induced subgap
  $\Delta/(k_B T)$ and chemical potential $\mu/(k_B T)$ with
  dimensionless parameter $K_{ph} = 3$ characterizing the phonon
  contribution to the thermal conductivity from the bulk states. The
  solid orange line represents the maximum $ZT$ at a fixed $\Delta$.}
\label{fig:ZT_with_kappa} 
\end{figure}

The phononic contribution to the thermal conductivity for the holey
sample can be characterized by the dimensionless parameter $K_{ph} =
2\kappa_{ph}{\cal D}h^2/(\tau k_B^3T^2)$. Here the phonon thermal
conductivity can be estimated to be $\kappa_{ph} \approx 1
\rm{Wm}^{-1}\rm{K}^{-1}$ (as for Bi$_2$Te$_3$) and the average
distance between the holes reaching ${\cal D}\sim 10$~nm. Taking the
relaxation time $\tau \approx 10^{-11}$ s, at room temperatures we
estimate $K_{ph} \sim 1$ for these rather conservative values of
parameters. In Fig.~\ref{fig:ZT_with_kappa} we show $ZT$ as a function
of the induced subgap $\Delta/(k_B T)$ and chemical potential
$\mu/(k_B T)$ for the phononic bulk contribution characterized by the
dimensionless parameter $K_{ph} = 3$. Although reduced considerably
from its pure 2D value, $ZT$ remains substantially larger than any
value so far achieved in these materials and can be tuned
significantly by changing the chemical potential, e. g., by gating the
system.

\begin{figure}
\includegraphics[width=1.0 \columnwidth]{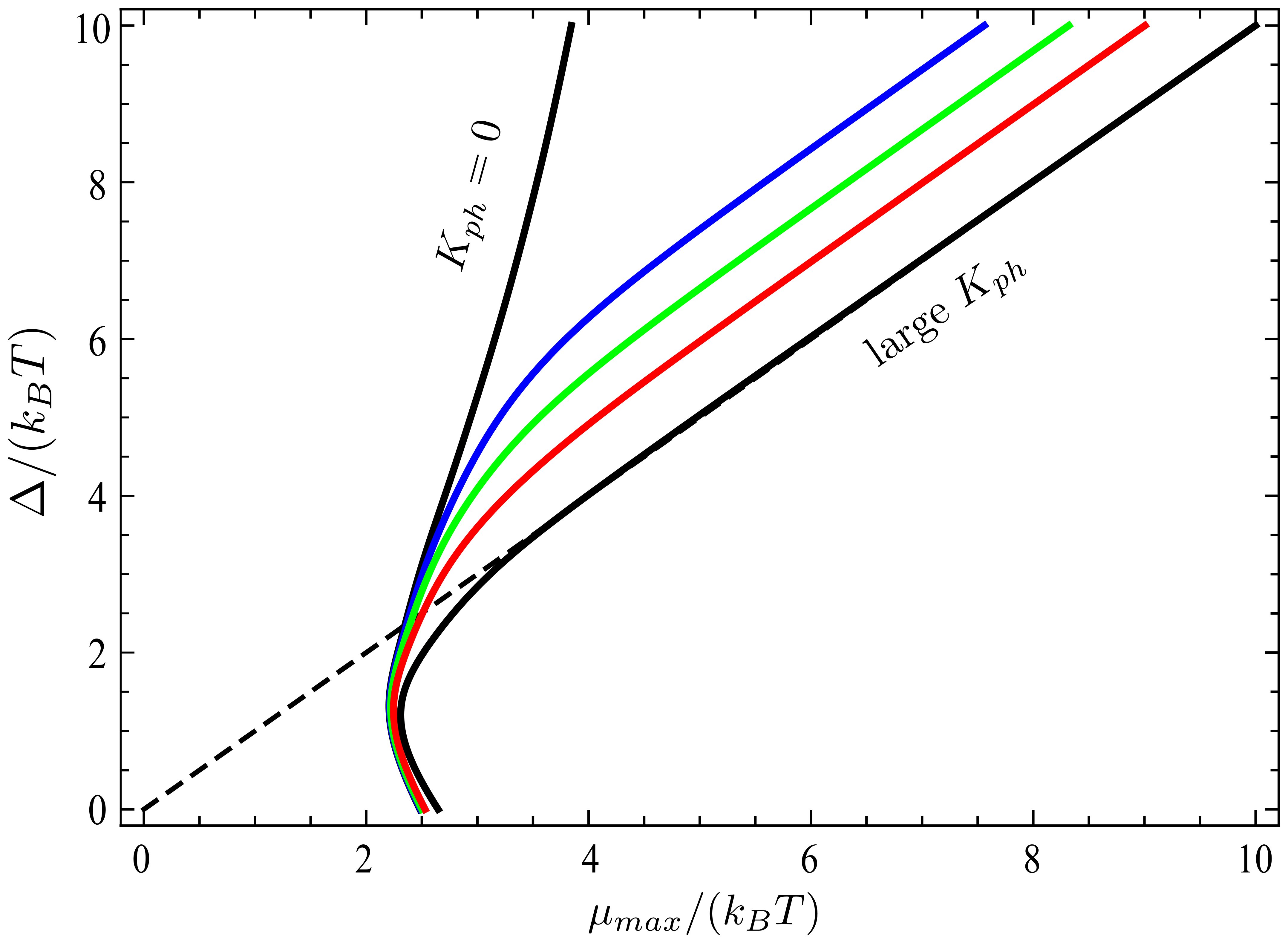} 
\caption{Induced subgap $\Delta/(k_B T)$ as a function of chemical
  potential $\mu_{max}/(k_B T)$ for the maximum value of $ZT$. The
  dimensionless parameter $K_{ph}$ characterizing the contribution of
  the phonon thermal conductivity is taken to be 0; 1; 3; 9; and
  90. The dashed line $\mu_{max} =\Delta$ is shown for comparison.}
\label{fig:mu_maxZT} 
\end{figure}

We also investigate the relation between the induced subgap
$\Delta/(k_B T)$ and the chemical potential $\mu_{max}/(k_B T)$ to
obtain the maximum value of $ZT$ for different phononic bulk
contributions. This result is shown in Fig.~\ref{fig:mu_maxZT}. For
small phonon thermal conductivity (small $K_{ph}$) the $ZT$ is maximal
when the chemical potential is placed within the subgap, $\mu <
\Delta$. In the limit of large $K_{ph}$ the maximum $ZT$ is reached
when the chemical potential and the gap match. This can be seen in
Fig.~\ref{fig:mu_maxZT}, where the line corresponding to $K_{ph}=90$
almost coincides with the dashed line $\Delta = \mu_{max}$ for rather
large $\mu_{max}/(k_B T)$.

\begin{figure}
\includegraphics[width=1.0 \columnwidth]{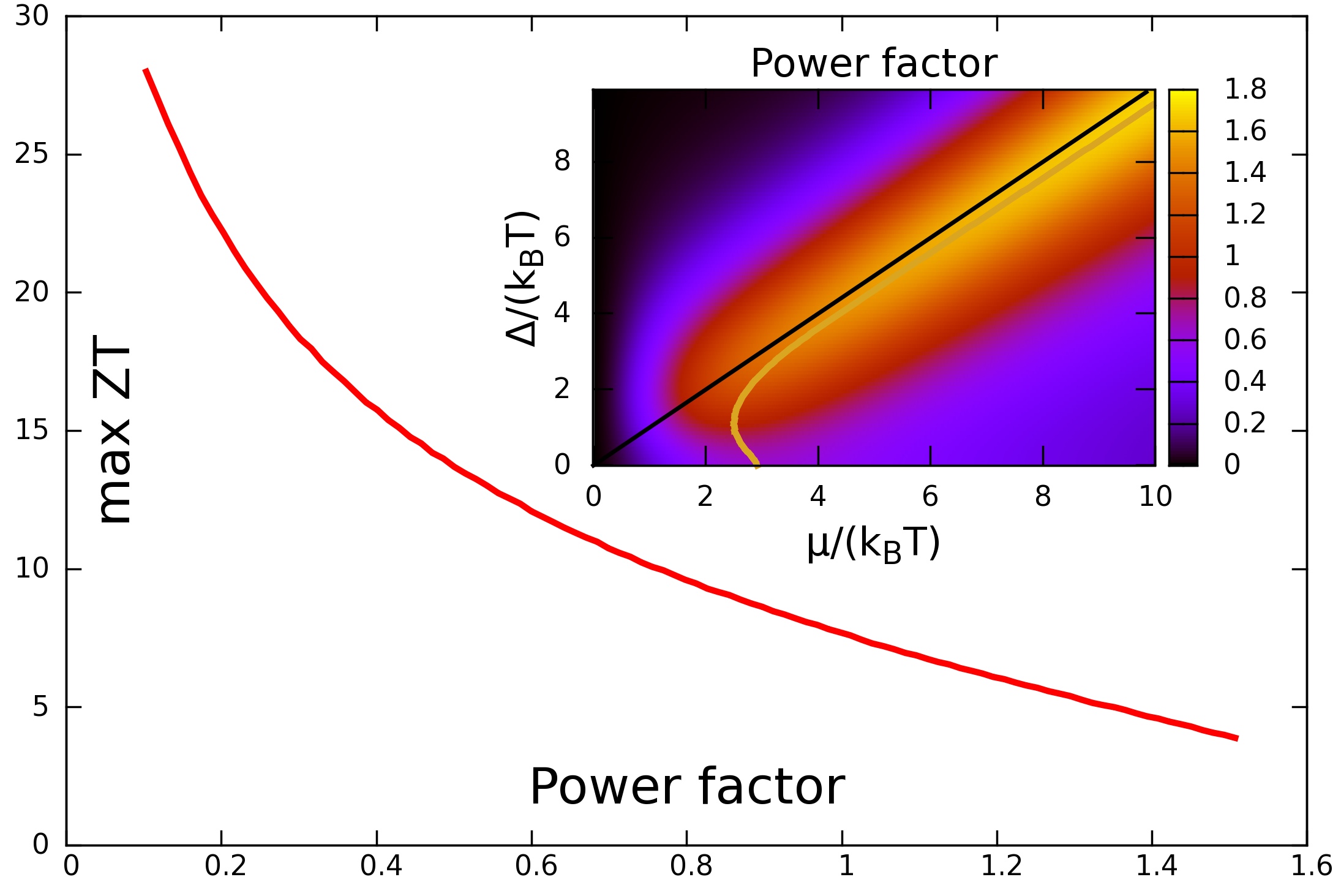} 
\caption{Maximum $ZT$ as a function of the dimensionless power factor,
  $L_1^2/L_0\propto \sigma S^2$. The inset shows the power factor as a
  function of the induced gap $\Delta/(k_B T)$ and chemical potential
  $\mu/(k_B T)$ measured in units of temperature.}
\label{fig:ZT_vs_Pout} 
\end{figure}

Finally, we point out that the large values of $ZT$ obtained in the
absence of phonon thermal conductivity shown in
Fig.~\ref{fig:ZT_gap_mu} is not the entire story. As at large
$\Delta/(k_B T)$ and $\mu/(k_B T)$ these thermoelectrics are indeed
very efficient but not effective -- the power factor $\sigma S^2$,
describing how much power one can produce, is small in this
limit. This is shown in Fig.~\ref{fig:ZT_vs_Pout} as a plot of the
maximal $ZT$ as a function of the power factor. The dimensionless
power output $L_1^2/L_0 \propto \sigma S^2$ as a function of the
induced subgap $\Delta/(k_B T)$ and chemical potential $\mu/(k_B T)$
is shown in the inset of Fig.~\ref{fig:ZT_vs_Pout}.

Within these materials, there are extensive competing mechanisms that
contribute to the thermal transport and thermoelectric efficiency. In
order to distinguish that a particular increase of $ZT$ is associated
with the topological protected states, we propose a simple measurement
of the transport coefficients with and without a magnetic field
applied, parallel to the transport direction, to increase the induced
gap of the surface states in the energy spectrum.  The effect of the
magnetic field will be more pronounced when the temperature is smaller
than the induced gap. For fields of $\sim 10$ Tesla a gap of the order
of several meV is expected in materials such as Bi$_2$Se$_3$
\cite{Zyuzin11, Analytis2010}.

It is important to note that our consideration is not limited to
periodic holey structures. It is also applicable to porous materials
with random size and location of the holes (pores) as shown in
Fig.~\ref{fig:porous}(a). In this case some of the surface states can
be situated too close to each other so that the back-scattering in
them will be increased, meanwhile the phonon part of thermal
conductivity will be reduced due to stronger trapping of phonons in
chaotic structures. Therefore, $ZT$ in porous materials can be
comparable to those with the periodically placed holes.

\section*{Methods}

To estimate the surface contribution of TI to the transport
coefficients $L_n$ we assume the bands to be Dirac-like with a subgap
$\Delta$ and use Boltzmann equation in the relaxation time
approximation,
\begin{equation}
L_{s,n} = - 2 \sum_{i}\int_{-\infty}^{\infty} 
\tau \left(\frac{\partial E_i}{\partial \hbar k}\right)^2  f'(E_i)  
(E_i-\mu)^{n} \frac{d^2 k}{4\pi^2}.
\label{Lsurf}
\end{equation}
Here the sum is over the upper and lower bands, $i=\pm 1$, and
$f' (E)=\partial f/\partial E$ with $f=1/(e^{(E-\mu)/(k_BT)}+1)$ being
the Fermi distribution function. Then taking relaxation time $\tau$ to
be independent of energy, we find
\begin{equation}
L_{s,n} = \frac{\tau (k_B T)^{1+n} }{2h^2} 
\!\int^{\infty}_{{\bar \Delta}}\!\!\! dx  
\frac{x^{2}-{\bar \Delta}^2}{x}
\left[ \frac{(x-{\bar \mu})^n}{\cosh^2 \frac{x-{\bar \mu}}{2}}
+\frac{(-x-{\bar \mu})^n}{\cosh^2 \frac{x+{\bar \mu}}{2}}\right],
\label{Ls}
\end{equation} 
where $h$ is Planck constant, ${\bar \Delta}=\Delta/(k_BT)$, and
${\bar \mu}=\mu/(k_BT)$.

To estimate the bulk contribution to $L_n$ we assume the bands to be
parabolic and find in the relaxation time approximation for the
conduction band,
\begin{equation}
L_{b,n} = - \tau \int_{\Delta_0}^{\infty}  D(E) 
\left(\frac{\partial E}{\partial \hbar k}\right)^2  f'(E)  (E-\mu)^{n}dE,
\label{Lbulk}
\end{equation}
where $D(E)$ is the density of states.  When the chemical potential is
far below the bottom of the conduction band, $(\Delta_0-\mu)/(k_B
T)\gg 1$, the contribution from the bulk to the transport coefficients
is exponentially suppressed, $L_{b,n}\propto e^{-(\Delta_0 -\mu)/(k_B
  T)}$, and can be safely neglected \cite{footnote4}. The same applies
for the valence band contribution when the chemical potential is far
away from the band edge.

Then the thermoelectric transport is dominated by the surface states
and the only sensible contribution from the bulk is to the phonon
thermal conductivity $\kappa_{ph}$, so that the figure of merit takes
the form
\begin{equation}
ZT=  
\frac{L_{s,1}^2}{L_{s,0}(L_{s,2}+{\cal D}\kappa_{ph}T) -L_{s,1}^2}.
\label{ZT_withKappa}
\end{equation}
This expression was used to obtain the results shown in
Figs.~\ref{fig:ZT_with_kappa} and~\ref{fig:mu_maxZT}.

\section*{Acknowledgments}

We thank M. Bakker, J. Heremans, C. Jaworski, J.~E. Moore, O. Mryasov,
and D. Pesin for insightful discussions. This work was supported by
NSF under Grant No. DMR-0547875, Grant No. 0757992, NSF-MRSEC
DMR-0820414, ONR-N000141110780, SWAN, and by the Welch Foundation
(A-1678).

\bibliography{thermoelectrictopological}

\begin{thebibliography}{29}%
\makeatletter
\providecommand \@ifxundefined [1]{%
 \@ifx{#1\undefined}
}%
\providecommand \@ifnum [1]{%
 \ifnum #1\expandafter \@firstoftwo
 \else \expandafter \@secondoftwo
 \fi
}%
\providecommand \@ifx [1]{%
 \ifx #1\expandafter \@firstoftwo
 \else \expandafter \@secondoftwo
 \fi
}%
\providecommand \natexlab [1]{#1}%
\providecommand \enquote  [1]{``#1''}%
\providecommand \bibnamefont  [1]{#1}%
\providecommand \bibfnamefont [1]{#1}%
\providecommand \citenamefont [1]{#1}%
\providecommand \href@noop [0]{\@secondoftwo}%
\providecommand \href [0]{\begingroup \@sanitize@url \@href}%
\providecommand \@href[1]{\@@startlink{#1}\@@href}%
\providecommand \@@href[1]{\endgroup#1\@@endlink}%
\providecommand \@sanitize@url [0]{\catcode `\\12\catcode `\$12\catcode
  `\&12\catcode `\#12\catcode `\^12\catcode `\_12\catcode `\%12\relax}%
\providecommand \@@startlink[1]{}%
\providecommand \@@endlink[0]{}%
\providecommand \url  [0]{\begingroup\@sanitize@url \@url }%
\providecommand \@url [1]{\endgroup\@href {#1}{\urlprefix }}%
\providecommand \urlprefix  [0]{URL }%
\providecommand \Eprint [0]{\href }%
\@ifxundefined \urlstyle {%
  \providecommand \doi  [0]{\begingroup \@sanitize@url \@doi}%
  \providecommand \@doi [1]{\endgroup \@@startlink {\doibase
  #1}doi:\discretionary {}{}{}#1\@@endlink }%
}{%
  \providecommand \doi  [0]{doi:\discretionary{}{}{}\begingroup
  \urlstyle{rm}\Url }%
}%
\providecommand \doibase [0]{http://dx.doi.org/}%
\providecommand \Doi [0]{\begingroup \@sanitize@url \@Doi }%
\providecommand \@Doi  [1]{\endgroup\@@startlink{\doibase#1}\@@Doi}%
\providecommand \@@Doi [1]{#1\@@endlink}%
\providecommand \selectlanguage [0]{\@gobble}%
\providecommand \bibinfo  [0]{\@secondoftwo}%
\providecommand \bibfield  [0]{\@secondoftwo}%
\providecommand \translation [1]{[#1]}%
\providecommand \BibitemOpen [0]{}%
\providecommand \bibitemStop [0]{}%
\providecommand \bibitemNoStop [0]{.\EOS\space}%
\providecommand \EOS [0]{\spacefactor3000\relax}%
\providecommand \BibitemShut  [1]{\csname bibitem#1\endcsname}%
\bibitem [{\citenamefont {Snyder}\ and\ \citenamefont
  {Toberer}(2008)}]{Snyder2008}%
  \BibitemOpen
  \bibfield  {author} {\bibinfo {author} {\bibfnamefont {G.~J.}\ \bibnamefont
  {Snyder}}\ and\ \bibinfo {author} {\bibfnamefont {E.~S.}\ \bibnamefont
  {Toberer}},\ }\Doi {10.1038/nmat2090} {\bibfield  {journal} {\bibinfo
  {journal} {Nat. Mater.},\ }\textbf {\bibinfo {volume} {7}},\ \bibinfo {pages}
  {105} (\bibinfo {year} {2008})}\BibitemShut {NoStop}%
\bibitem [{\citenamefont {Tritt}(1999)}]{Tritt99}%
  \BibitemOpen
  \bibfield  {author} {\bibinfo {author} {\bibfnamefont {T.~M.}\ \bibnamefont
  {Tritt}},\ }\Doi {10.1126/science.283.5403.804} {\bibfield  {journal}
  {\bibinfo  {journal} {Science},\ }\textbf {\bibinfo {volume} {283}},\
  \bibinfo {pages} {804} (\bibinfo {year} {1999})}\BibitemShut {NoStop}%
\bibitem [{\citenamefont {Hicks}\ and\ \citenamefont
  {Dresselhaus}(1993)}]{Dresselhaus93}%
  \BibitemOpen
  \bibfield  {author} {\bibinfo {author} {\bibfnamefont {L.~D.}\ \bibnamefont
  {Hicks}}\ and\ \bibinfo {author} {\bibfnamefont {M.~S.}\ \bibnamefont
  {Dresselhaus}},\ }\Doi {10.1103/PhysRevB.47.12727} {\bibfield  {journal}
  {\bibinfo  {journal} {Phys. Rev. B},\ }\textbf {\bibinfo {volume} {47}},\
  \bibinfo {pages} {12727} (\bibinfo {year} {1993})}\BibitemShut {NoStop}%
\bibitem [{\citenamefont {Mukerjee}\ and\ \citenamefont
  {Moore}(2007)}]{mukerjee07}%
  \BibitemOpen
  \bibfield  {author} {\bibinfo {author} {\bibfnamefont {S.}~\bibnamefont
  {Mukerjee}}\ and\ \bibinfo {author} {\bibfnamefont {J.~E.}\ \bibnamefont
  {Moore}},\ }\Doi {10.1063/1.2712775} {\bibfield  {journal} {\bibinfo
  {journal} {Appl. Phys. Lett.},\ }\textbf {\bibinfo {volume} {90}},\ \bibinfo
  {eid} {112107} (\bibinfo {year} {2007})}\BibitemShut {NoStop}%
\bibitem [{\citenamefont {Markussen}\ \emph {et~al.}(2009)\citenamefont
  {Markussen}, \citenamefont {Jauho},\ and\ \citenamefont
  {Brandbyge}}]{Markussen09}%
  \BibitemOpen
  \bibfield  {author} {\bibinfo {author} {\bibfnamefont {T.}~\bibnamefont
  {Markussen}}, \bibinfo {author} {\bibfnamefont {A.-P.}\ \bibnamefont
  {Jauho}}, \ and\ \bibinfo {author} {\bibfnamefont {M.}~\bibnamefont
  {Brandbyge}},\ }\Doi {10.1103/PhysRevLett.103.055502} {\bibfield  {journal}
  {\bibinfo  {journal} {Phys. Rev. Lett.},\ }\textbf {\bibinfo {volume}
  {103}},\ \bibinfo {pages} {055502} (\bibinfo {year} {2009})}\BibitemShut
  {NoStop}%
\bibitem [{\citenamefont {Takahashi}\ and\ \citenamefont
  {Murakami}(2010)}]{Murakami2010}%
  \BibitemOpen
  \bibfield  {author} {\bibinfo {author} {\bibfnamefont {R.}~\bibnamefont
  {Takahashi}}\ and\ \bibinfo {author} {\bibfnamefont {S.}~\bibnamefont
  {Murakami}},\ }\Doi {10.1103/PhysRevB.81.161302} {\bibfield  {journal}
  {\bibinfo  {journal} {Phys. Rev. B},\ }\textbf {\bibinfo {volume} {81}},\
  \bibinfo {pages} {161302} (\bibinfo {year} {2010})}\BibitemShut {NoStop}%
\bibitem [{\citenamefont {Ghaemi}\ \emph {et~al.}(2010)\citenamefont {Ghaemi},
  \citenamefont {Mong},\ and\ \citenamefont {Moore}}]{Ghaemi10}%
  \BibitemOpen
  \bibfield  {author} {\bibinfo {author} {\bibfnamefont {P.}~\bibnamefont
  {Ghaemi}}, \bibinfo {author} {\bibfnamefont {R.~S.~K.}\ \bibnamefont {Mong}},
  \ and\ \bibinfo {author} {\bibfnamefont {J.~E.}\ \bibnamefont {Moore}},\
  }\Doi {10.1103/PhysRevLett.105.166603} {\bibfield  {journal} {\bibinfo
  {journal} {Phys. Rev. Lett.},\ }\textbf {\bibinfo {volume} {105}},\ \bibinfo
  {pages} {166603} (\bibinfo {year} {2010})}\BibitemShut {NoStop}%
\bibitem [{\citenamefont {Tretiakov}\ \emph {et~al.}(2010)\citenamefont
  {Tretiakov}, \citenamefont {Abanov}, \citenamefont {Murakami},\ and\
  \citenamefont {Sinova}}]{TretiakovAPL10}%
  \BibitemOpen
  \bibfield  {author} {\bibinfo {author} {\bibfnamefont {O.~A.}\ \bibnamefont
  {Tretiakov}}, \bibinfo {author} {\bibfnamefont {A.}~\bibnamefont {Abanov}},
  \bibinfo {author} {\bibfnamefont {S.}~\bibnamefont {Murakami}}, \ and\
  \bibinfo {author} {\bibfnamefont {J.}~\bibnamefont {Sinova}},\ }\Doi
  {10.1063/1.3481382} {\bibfield  {journal} {\bibinfo  {journal} {Appl. Phys.
  Lett.},\ }\textbf {\bibinfo {volume} {97}},\ \bibinfo {eid} {073108}
  (\bibinfo {year} {2010})}\BibitemShut {NoStop}%
\bibitem [{\citenamefont {Lyeo}\ \emph {et~al.}(2004)\citenamefont {Lyeo},
  \citenamefont {Khajetoorians}, \citenamefont {Shi}, \citenamefont {Pipe},
  \citenamefont {Ram}, \citenamefont {Shakouri},\ and\ \citenamefont
  {Shih}}]{Ho-KiLyeo04}%
  \BibitemOpen
  \bibfield  {author} {\bibinfo {author} {\bibfnamefont {H.-K.}\ \bibnamefont
  {Lyeo}}, \bibinfo {author} {\bibfnamefont {A.~A.}\ \bibnamefont
  {Khajetoorians}}, \bibinfo {author} {\bibfnamefont {L.}~\bibnamefont {Shi}},
  \bibinfo {author} {\bibfnamefont {K.~P.}\ \bibnamefont {Pipe}}, \bibinfo
  {author} {\bibfnamefont {R.~J.}\ \bibnamefont {Ram}}, \bibinfo {author}
  {\bibfnamefont {A.}~\bibnamefont {Shakouri}}, \ and\ \bibinfo {author}
  {\bibfnamefont {C.~K.}\ \bibnamefont {Shih}},\ }\Doi
  {10.1126/science.1091600} {\bibfield  {journal} {\bibinfo  {journal}
  {Science},\ }\textbf {\bibinfo {volume} {303}},\ \bibinfo {pages} {816}
  (\bibinfo {year} {2004})}\BibitemShut {NoStop}%
\bibitem [{\citenamefont {Venkatasubramanian}\ \emph
  {et~al.}(2001)\citenamefont {Venkatasubramanian}, \citenamefont {Siivola},
  \citenamefont {Colpitts},\ and\ \citenamefont
  {{O'Quinn}}}]{venkatasubramanian01}%
  \BibitemOpen
  \bibfield  {author} {\bibinfo {author} {\bibfnamefont {R.}~\bibnamefont
  {Venkatasubramanian}}, \bibinfo {author} {\bibfnamefont {E.}~\bibnamefont
  {Siivola}}, \bibinfo {author} {\bibfnamefont {T.}~\bibnamefont {Colpitts}}, \
  and\ \bibinfo {author} {\bibfnamefont {B.}~\bibnamefont {{O'Quinn}}},\ }\Doi
  {10.1038/35098012} {\bibfield  {journal} {\bibinfo  {journal} {Nature},\
  }\textbf {\bibinfo {volume} {413}},\ \bibinfo {pages} {597} (\bibinfo {year}
  {2001})}\BibitemShut {NoStop}%
\bibitem [{\citenamefont {Zhang}\ \emph {et~al.}(2010)\citenamefont {Zhang},
  \citenamefont {Hapenciuc}, \citenamefont {Castillo}, \citenamefont
  {Borca-Tasciuc}, \citenamefont {Mehta}, \citenamefont {Karthik},\ and\
  \citenamefont {Ramanath}}]{zhang:062107}%
  \BibitemOpen
  \bibfield  {author} {\bibinfo {author} {\bibfnamefont {Y.}~\bibnamefont
  {Zhang}}, \bibinfo {author} {\bibfnamefont {C.~L.}\ \bibnamefont
  {Hapenciuc}}, \bibinfo {author} {\bibfnamefont {E.~E.}\ \bibnamefont
  {Castillo}}, \bibinfo {author} {\bibfnamefont {T.}~\bibnamefont
  {Borca-Tasciuc}}, \bibinfo {author} {\bibfnamefont {R.~J.}\ \bibnamefont
  {Mehta}}, \bibinfo {author} {\bibfnamefont {C.}~\bibnamefont {Karthik}}, \
  and\ \bibinfo {author} {\bibfnamefont {G.}~\bibnamefont {Ramanath}},\ }\Doi
  {10.1063/1.3300826} {\bibfield  {journal} {\bibinfo  {journal} {Appl. Phys.
  Lett.},\ }\textbf {\bibinfo {volume} {96}},\ \bibinfo {eid} {062107}
  (\bibinfo {year} {2010})}\BibitemShut {NoStop}%
\bibitem [{\citenamefont {Teweldebrhan}\ \emph
  {et~al.}(2010){\natexlab{a}}\citenamefont {Teweldebrhan}, \citenamefont
  {Goyal}, \citenamefont {Rahman},\ and\ \citenamefont
  {Balandin}}]{TeweldebrhanAPL10}%
  \BibitemOpen
  \bibfield  {author} {\bibinfo {author} {\bibfnamefont {D.}~\bibnamefont
  {Teweldebrhan}}, \bibinfo {author} {\bibfnamefont {V.}~\bibnamefont {Goyal}},
  \bibinfo {author} {\bibfnamefont {M.}~\bibnamefont {Rahman}}, \ and\ \bibinfo
  {author} {\bibfnamefont {A.~A.}\ \bibnamefont {Balandin}},\ }\Doi
  {10.1063/1.3280078} {\bibfield  {journal} {\bibinfo  {journal} {Appl. Phys.
  Lett.},\ }\textbf {\bibinfo {volume} {96}},\ \bibinfo {eid} {053107}
  (\bibinfo {year} {2010}{\natexlab{a}})}\BibitemShut {NoStop}%
\bibitem [{\citenamefont {Teweldebrhan}\ \emph
  {et~al.}(2010){\natexlab{b}}\citenamefont {Teweldebrhan}, \citenamefont
  {Goyal},\ and\ \citenamefont {Balandin}}]{Teweldebrhan10}%
  \BibitemOpen
  \bibfield  {author} {\bibinfo {author} {\bibfnamefont {D.}~\bibnamefont
  {Teweldebrhan}}, \bibinfo {author} {\bibfnamefont {V.}~\bibnamefont {Goyal}},
  \ and\ \bibinfo {author} {\bibfnamefont {A.~A.}\ \bibnamefont {Balandin}},\
  }\Doi {10.1021/nl903590b} {\bibfield  {journal} {\bibinfo  {journal} {Nano
  Lett.},\ }\textbf {\bibinfo {volume} {10}},\ \bibinfo {pages} {1209}
  (\bibinfo {year} {2010}{\natexlab{b}})}\BibitemShut {NoStop}%
\bibitem [{\citenamefont {Dubi}\ and\ \citenamefont
  {Di~Ventra}(2011)}]{DiVentra11}%
  \BibitemOpen
  \bibfield  {author} {\bibinfo {author} {\bibfnamefont {Y.}~\bibnamefont
  {Dubi}}\ and\ \bibinfo {author} {\bibfnamefont {M.}~\bibnamefont
  {Di~Ventra}},\ }\Doi {10.1103/RevModPhys.83.131} {\bibfield  {journal}
  {\bibinfo  {journal} {Rev. Mod. Phys.},\ }\textbf {\bibinfo {volume} {83}},\
  \bibinfo {pages} {131} (\bibinfo {year} {2011})}\BibitemShut {NoStop}%
\bibitem [{\citenamefont {Qi}\ and\ \citenamefont
  {Zhang}(2010)}]{TI_physics_today}%
  \BibitemOpen
  \bibfield  {author} {\bibinfo {author} {\bibfnamefont {X.-L.}\ \bibnamefont
  {Qi}}\ and\ \bibinfo {author} {\bibfnamefont {S.-C.}\ \bibnamefont {Zhang}},\
  }\href@noop {} {\bibfield  {journal} {\bibinfo  {journal} {Physics Today},\
  }\textbf {\bibinfo {volume} {63}},\ \bibinfo {pages} {33} (\bibinfo {year}
  {2010})}\BibitemShut {NoStop}%
\bibitem [{\citenamefont {Fu}\ \emph {et~al.}(2007)\citenamefont {Fu},
  \citenamefont {Kane},\ and\ \citenamefont {Mele}}]{Fu07}%
  \BibitemOpen
  \bibfield  {author} {\bibinfo {author} {\bibfnamefont {L.}~\bibnamefont
  {Fu}}, \bibinfo {author} {\bibfnamefont {C.~L.}\ \bibnamefont {Kane}}, \ and\
  \bibinfo {author} {\bibfnamefont {E.~J.}\ \bibnamefont {Mele}},\ }\Doi
  {10.1103/PhysRevLett.98.106803} {\bibfield  {journal} {\bibinfo  {journal}
  {Phys. Rev. Lett.},\ }\textbf {\bibinfo {volume} {98}},\ \bibinfo {pages}
  {106803} (\bibinfo {year} {2007})}\BibitemShut {NoStop}%
\bibitem [{\citenamefont {Hsieh}\ \emph {et~al.}(2008)\citenamefont {Hsieh},
  \citenamefont {Qian}, \citenamefont {Wray}, \citenamefont {Xia},
  \citenamefont {Hor}, \citenamefont {Cava},\ and\ \citenamefont
  {Hasan}}]{Hsieh2008}%
  \BibitemOpen
  \bibfield  {author} {\bibinfo {author} {\bibfnamefont {D.}~\bibnamefont
  {Hsieh}}, \bibinfo {author} {\bibfnamefont {D.}~\bibnamefont {Qian}},
  \bibinfo {author} {\bibfnamefont {L.}~\bibnamefont {Wray}}, \bibinfo {author}
  {\bibfnamefont {Y.}~\bibnamefont {Xia}}, \bibinfo {author} {\bibfnamefont
  {Y.~S.}\ \bibnamefont {Hor}}, \bibinfo {author} {\bibfnamefont {R.~J.}\
  \bibnamefont {Cava}}, \ and\ \bibinfo {author} {\bibfnamefont {M.~Z.}\
  \bibnamefont {Hasan}},\ }\Doi {10.1038/nature06843} {\bibfield  {journal}
  {\bibinfo  {journal} {Nature},\ }\textbf {\bibinfo {volume} {452}},\ \bibinfo
  {pages} {970} (\bibinfo {year} {2008})}\BibitemShut {NoStop}%
\bibitem [{\citenamefont {Chen}\ \emph {et~al.}(2009)\citenamefont {Chen},
  \citenamefont {Analytis}, \citenamefont {Chu}, \citenamefont {Liu},
  \citenamefont {Mo}, \citenamefont {Qi}, \citenamefont {Zhang}, \citenamefont
  {Lu}, \citenamefont {Dai}, \citenamefont {Fang}, \citenamefont {Zhang},
  \citenamefont {Fisher}, \citenamefont {Hussain},\ and\ \citenamefont
  {Shen}}]{Chen2009}%
  \BibitemOpen
  \bibfield  {author} {\bibinfo {author} {\bibfnamefont {Y.~L.}\ \bibnamefont
  {Chen}}, \bibinfo {author} {\bibfnamefont {J.~G.}\ \bibnamefont {Analytis}},
  \bibinfo {author} {\bibfnamefont {J.-H.}\ \bibnamefont {Chu}}, \bibinfo
  {author} {\bibfnamefont {Z.~K.}\ \bibnamefont {Liu}}, \bibinfo {author}
  {\bibfnamefont {S.-K.}\ \bibnamefont {Mo}}, \bibinfo {author} {\bibfnamefont
  {X.~L.}\ \bibnamefont {Qi}}, \bibinfo {author} {\bibfnamefont {H.~J.}\
  \bibnamefont {Zhang}}, \bibinfo {author} {\bibfnamefont {D.~H.}\ \bibnamefont
  {Lu}}, \bibinfo {author} {\bibfnamefont {X.}~\bibnamefont {Dai}}, \bibinfo
  {author} {\bibfnamefont {Z.}~\bibnamefont {Fang}}, \bibinfo {author}
  {\bibfnamefont {S.~C.}\ \bibnamefont {Zhang}}, \bibinfo {author}
  {\bibfnamefont {I.~R.}\ \bibnamefont {Fisher}}, \bibinfo {author}
  {\bibfnamefont {Z.}~\bibnamefont {Hussain}}, \ and\ \bibinfo {author}
  {\bibfnamefont {Z.-X.}\ \bibnamefont {Shen}},\ }\Doi
  {10.1126/science.1173034} {\bibfield  {journal} {\bibinfo  {journal}
  {Science},\ }\textbf {\bibinfo {volume} {325}},\ \bibinfo {pages} {178}
  (\bibinfo {year} {2009})}\BibitemShut {NoStop}%
\bibitem [{\citenamefont {Poudel}\ \emph {et~al.}(2008)\citenamefont {Poudel},
  \citenamefont {Hao}, \citenamefont {Ma}, \citenamefont {Lan}, \citenamefont
  {Minnich}, \citenamefont {Yu}, \citenamefont {Yan}, \citenamefont {Wang},
  \citenamefont {Muto}, \citenamefont {Vashaee}, \citenamefont {Chen},
  \citenamefont {Liu}, \citenamefont {Dresselhaus}, \citenamefont {Chen},\ and\
  \citenamefont {Ren}}]{Poudel2008}%
  \BibitemOpen
  \bibfield  {author} {\bibinfo {author} {\bibfnamefont {B.}~\bibnamefont
  {Poudel}}, \bibinfo {author} {\bibfnamefont {Q.}~\bibnamefont {Hao}},
  \bibinfo {author} {\bibfnamefont {Y.}~\bibnamefont {Ma}}, \bibinfo {author}
  {\bibfnamefont {Y.}~\bibnamefont {Lan}}, \bibinfo {author} {\bibfnamefont
  {A.}~\bibnamefont {Minnich}}, \bibinfo {author} {\bibfnamefont
  {B.}~\bibnamefont {Yu}}, \bibinfo {author} {\bibfnamefont {X.}~\bibnamefont
  {Yan}}, \bibinfo {author} {\bibfnamefont {D.}~\bibnamefont {Wang}}, \bibinfo
  {author} {\bibfnamefont {A.}~\bibnamefont {Muto}}, \bibinfo {author}
  {\bibfnamefont {D.}~\bibnamefont {Vashaee}}, \bibinfo {author} {\bibfnamefont
  {X.}~\bibnamefont {Chen}}, \bibinfo {author} {\bibfnamefont {J.}~\bibnamefont
  {Liu}}, \bibinfo {author} {\bibfnamefont {M.~S.}\ \bibnamefont
  {Dresselhaus}}, \bibinfo {author} {\bibfnamefont {G.}~\bibnamefont {Chen}}, \
  and\ \bibinfo {author} {\bibfnamefont {Z.}~\bibnamefont {Ren}},\ }\Doi
  {10.1126/science.1156446} {\bibfield  {journal} {\bibinfo  {journal}
  {Science},\ }\textbf {\bibinfo {volume} {320}},\ \bibinfo {pages} {634}
  (\bibinfo {year} {2008})}\BibitemShut {NoStop}%
\bibitem [{\citenamefont {Tang}\ \emph {et~al.}(2010)\citenamefont {Tang},
  \citenamefont {Wang}, \citenamefont {Lee}, \citenamefont {Fardy},
  \citenamefont {Huo}, \citenamefont {Russell},\ and\ \citenamefont
  {Yang}}]{Tang10}%
  \BibitemOpen
  \bibfield  {author} {\bibinfo {author} {\bibfnamefont {J.}~\bibnamefont
  {Tang}}, \bibinfo {author} {\bibfnamefont {H.-T.}\ \bibnamefont {Wang}},
  \bibinfo {author} {\bibfnamefont {D.~H.}\ \bibnamefont {Lee}}, \bibinfo
  {author} {\bibfnamefont {M.}~\bibnamefont {Fardy}}, \bibinfo {author}
  {\bibfnamefont {Z.}~\bibnamefont {Huo}}, \bibinfo {author} {\bibfnamefont
  {T.~P.}\ \bibnamefont {Russell}}, \ and\ \bibinfo {author} {\bibfnamefont
  {P.}~\bibnamefont {Yang}},\ }\Doi {10.1021/nl102931z} {\bibfield  {journal}
  {\bibinfo  {journal} {Nano Lett.},\ }\textbf {\bibinfo {volume} {10}},\
  \bibinfo {pages} {4279} (\bibinfo {year} {2010})}\BibitemShut {NoStop}%
\bibitem [{\citenamefont {Park}\ \emph {et~al.}(2010)\citenamefont {Park},
  \citenamefont {Heremans}, \citenamefont {Scarola},\ and\ \citenamefont
  {Minic}}]{ParkPRL10}%
  \BibitemOpen
  \bibfield  {author} {\bibinfo {author} {\bibfnamefont {K.}~\bibnamefont
  {Park}}, \bibinfo {author} {\bibfnamefont {J.~J.}\ \bibnamefont {Heremans}},
  \bibinfo {author} {\bibfnamefont {V.~W.}\ \bibnamefont {Scarola}}, \ and\
  \bibinfo {author} {\bibfnamefont {D.}~\bibnamefont {Minic}},\ }\Doi
  {10.1103/PhysRevLett.105.186801} {\bibfield  {journal} {\bibinfo  {journal}
  {Phys. Rev. Lett.},\ }\textbf {\bibinfo {volume} {105}},\ \bibinfo {pages}
  {186801} (\bibinfo {year} {2010})}\BibitemShut {NoStop}%
\bibitem [{Not()}]{Note1}%
  \BibitemOpen
  \href@noop {} {}\bibinfo {note} {We consider $v$ to be energy independent in
  a wide range of energies}\BibitemShut {NoStop}%
\bibitem [{\citenamefont {Linder}\ \emph {et~al.}(2009)\citenamefont {Linder},
  \citenamefont {Yokoyama},\ and\ \citenamefont {Sudb\o{}}}]{Linder09}%
  \BibitemOpen
  \bibfield  {author} {\bibinfo {author} {\bibfnamefont {J.}~\bibnamefont
  {Linder}}, \bibinfo {author} {\bibfnamefont {T.}~\bibnamefont {Yokoyama}}, \
  and\ \bibinfo {author} {\bibfnamefont {A.}~\bibnamefont {Sudb\o{}}},\ }\Doi
  {10.1103/PhysRevB.80.205401} {\bibfield  {journal} {\bibinfo  {journal}
  {Phys. Rev. B},\ }\textbf {\bibinfo {volume} {80}},\ \bibinfo {pages}
  {205401} (\bibinfo {year} {2009})}\BibitemShut {NoStop}%
\bibitem [{\citenamefont {Lu}\ \emph {et~al.}(2010)\citenamefont {Lu},
  \citenamefont {Shan}, \citenamefont {Yao}, \citenamefont {Niu},\ and\
  \citenamefont {Shen}}]{LuPRB10}%
  \BibitemOpen
  \bibfield  {author} {\bibinfo {author} {\bibfnamefont {H.-Z.}\ \bibnamefont
  {Lu}}, \bibinfo {author} {\bibfnamefont {W.-Y.}\ \bibnamefont {Shan}},
  \bibinfo {author} {\bibfnamefont {W.}~\bibnamefont {Yao}}, \bibinfo {author}
  {\bibfnamefont {Q.}~\bibnamefont {Niu}}, \ and\ \bibinfo {author}
  {\bibfnamefont {S.-Q.}\ \bibnamefont {Shen}},\ }\Doi
  {10.1103/PhysRevB.81.115407} {\bibfield  {journal} {\bibinfo  {journal}
  {Phys. Rev. B},\ }\textbf {\bibinfo {volume} {81}},\ \bibinfo {pages}
  {115407} (\bibinfo {year} {2010})}\BibitemShut {NoStop}%
\bibitem [{\citenamefont {Ashcroft}\ and\ \citenamefont
  {Mermin}(1976)}]{AshcroftMermin}%
  \BibitemOpen
  \bibfield  {author} {\bibinfo {author} {\bibfnamefont {N.~W.}\ \bibnamefont
  {Ashcroft}}\ and\ \bibinfo {author} {\bibfnamefont {N.~D.}\ \bibnamefont
  {Mermin}},\ }\href@noop {} {\emph {\bibinfo {title} {Solid State Physics}}}\
  (\bibinfo  {publisher} {Sauders College Publishing},\ \bibinfo {address}
  {Fort Worth},\ \bibinfo {year} {1976})\BibitemShut {NoStop}%
\bibitem [{\citenamefont {Marder}(1976)}]{marder}%
  \BibitemOpen
  \bibfield  {author} {\bibinfo {author} {\bibfnamefont {M.~P.}\ \bibnamefont
  {Marder}},\ }\href@noop {} {\emph {\bibinfo {title} {Condensed Matter
  Physics}}}\ (\bibinfo  {publisher} {John Wiley \& Sons, Inc.},\ \bibinfo
  {address} {New York},\ \bibinfo {year} {1976})\BibitemShut {NoStop}%
\bibitem [{\citenamefont {Zyuzin}\ and\ \citenamefont
  {Burkov}(2011)}]{Zyuzin11}%
  \BibitemOpen
  \bibfield  {author} {\bibinfo {author} {\bibfnamefont {A.~A.}\ \bibnamefont
  {Zyuzin}}\ and\ \bibinfo {author} {\bibfnamefont {A.~A.}\ \bibnamefont
  {Burkov}},\ }\Doi {10.1103/PhysRevB.83.195413} {\bibfield  {journal}
  {\bibinfo  {journal} {Phys. Rev. B},\ }\textbf {\bibinfo {volume} {83}},\
  \bibinfo {pages} {195413} (\bibinfo {year} {2011})}\BibitemShut {NoStop}%
\bibitem [{\citenamefont {Analytis}\ \emph {et~al.}(2010)\citenamefont
  {Analytis}, \citenamefont {{McDonald}}, \citenamefont {Riggs}, \citenamefont
  {Chu}, \citenamefont {Boebinger},\ and\ \citenamefont
  {Fisher}}]{Analytis2010}%
  \BibitemOpen
  \bibfield  {author} {\bibinfo {author} {\bibfnamefont {J.~G.}\ \bibnamefont
  {Analytis}}, \bibinfo {author} {\bibfnamefont {R.~D.}\ \bibnamefont
  {{McDonald}}}, \bibinfo {author} {\bibfnamefont {S.~C.}\ \bibnamefont
  {Riggs}}, \bibinfo {author} {\bibfnamefont {J.}~\bibnamefont {Chu}}, \bibinfo
  {author} {\bibfnamefont {G.~S.}\ \bibnamefont {Boebinger}}, \ and\ \bibinfo
  {author} {\bibfnamefont {I.~R.}\ \bibnamefont {Fisher}},\ }\Doi
  {10.1038/nphys1861} {\bibfield  {journal} {\bibinfo  {journal} {Nat. Phys.},\
  }\textbf {\bibinfo {volume} {6}},\ \bibinfo {pages} {960} (\bibinfo {year}
  {2010})}\BibitemShut {NoStop}%
\bibitem [{foo()}]{footnote4}%
  \BibitemOpen
  \href@noop {} {}\bibinfo {howpublished} {We note that taking into account the
  fact that the extended and localized states in the bulk are separated by the
  mobility edge $E_m$ which is measured from the bottom of the band changes
  $\Delta_0$ to $\Delta_0 - E_m$ in these estimates and only makes the
  inequalities to be even stronger}\BibitemShut {NoStop}%
\end{thebibliography}%

\end{document}